\documentclass[showpacs,amsmath,twocolumn,superscriptaddress]{revtex4}
\usepackage{graphicx}
\usepackage[caption=false]{subfig}
\usepackage[all]{xy}
\usepackage{amsmath}
\usepackage{amssymb}
\usepackage{enumerate}
\newcommand{\be}{\begin{equation}}
\newcommand{\ee}{\end{equation}}
\newcommand{\ben}{\begin{eqnarray}}
\newcommand{\een}{\end{eqnarray}}
\newcommand{\bes}{\begin{subequations}}
\newcommand{\ees}{\end{subequations}}
\newcommand{\bb}{\bibitem}

\newcommand{\nn}{\nonumber\\}
\newcommand{\bfi}{\begin{figure}}
\newcommand{\efi}{\end{figure}}
\newcommand{\bc}{\begin{center}}
\newcommand{\ec}{\end{center}}
\newcommand{\sech}{\mbox{sech}}

\begin{document}
\title{Braneworld solutions for F(R) models with non-constant curvature}

\author{D. Bazeia} 
\affiliation{Departamento de F\'\i sica, Universidade Federal da Para\'\i ba, 58051-970 Jo\~ao Pessoa, PB, Brazil}
\affiliation{Departamento de F\'\i sica, Universidade Federal de Campina Grande, 58109-970 Campina Grande, PB, Brazil}
%\email{bazeia@fisica.ufpb.br}

\author{ A. S. Lob\~ao Jr}
\affiliation{Departamento de F\'\i sica, Universidade Federal da Para\'\i ba, 58051-970 Jo\~ao Pessoa, PB, Brazil}

\author{R. Menezes}
\affiliation{Departamento de Ci\^encias Exatas, Universidade Federal da Para\'\i ba, 58297-000 Rio Tinto, PB, Brazil.}
\affiliation{Departamento de F\'\i sica, Universidade Federal de Campina Grande, 58109-970 Campina Grande, PB, Brazil}
%\email{rmenenezes@dce.ufpb.br}

\author{A. Yu. Petrov}

%\email{petrov@fisica.ufpb.br}
\affiliation{Departamento de F\'\i sica, Universidade Federal da Para\'\i ba, 58051-970 Jo\~ao Pessoa, PB, Brazil}

\author{A. J. da Silva}
\affiliation{Instituto de F\'\i sica Universidade de S\~ao Paulo, 05314-970 S\~ao Paulo SP, Brazil}

\begin{abstract}
This work deal with braneworld scenarios with generalized gravity. We investigate models where the potential of the scalar field is polynomial or nonpolynomial. We obtain exact and approximated solutions for the scalar field, warp factor and energy density, in the complex scenario with no restriction on the scalar curvature. In particular, we describe the case where the brane may split, engendering internal structure, with the splitting caused by the same parameter that controls deviation from standard gravity.

\end{abstract}

\pacs{11.25.Uv}

\maketitle

%%%%%%%%%%%%%%%%%%%%%%%%%%
\section{introduction}

The Randall-Sundrum (RS) theory proposes a description of the Universe with a five-dimensional ($5D$) spacetime of the anti de Sitter ($AdS$) type, with a single extra spatial dimension of infinite extent. In this scenario, our Universe evolves as a brane embedded in a $5D$ bulk where gravity can propagates, but with the other fundamental interactions only propagating on the brane. The RS and other \cite{Nima,GW,AH,F,G,S1,S2,D,C} braneworld scenarios are of current interest because they help us to better understand the cosmological constant and mass hierarchy problems \cite{RS,Nima,GW,AH,F,G,S1,S2,D,C}.

The study of branes requires solving Einstein equations and modified Einstein equations in the case of modified gravity, in an $AdS_5$ spacetime; see, e.g., Refs.~\cite{B,AB,BMP}. Due to the interest in modified gravity (see e.g., Ref.~{\cite{mg}), we have investigated braneworld scenarios under the substitution  $R\to F(R)$ in the action that describes the brane; see, e.g., \cite{AB,BMP} .
An important tool for such calculations is the first-order formalism, which allows to reduce the order of the equations of motion and is important for generalized models, where the degree of complexity of the equations of motion is very high \cite{FO}.  

In the case of modified gravity, the investigations presented in Refs.~\cite{AB,BMP} were based on the assumption that the scalar curvature is constant. In the current work, however, we abandon this restriction and solve the brane equations in the more general situation, for non-constant scalar curvature. Due to the complexity of the brane equations, we work with models described by $F(R)\!=\!R +\alpha R^2$, and we consider the case of a single real scalar field. To obtain a larger family of solutions, we follow two distinct routes: firstly, we deal with the brane equations via an exact procedure; in the second case, we develop an approximation scheme, in which we consider $\alpha$ small, working up to the first-order in $\alpha$. We study a single scalar field, but we work with models having polynomial interaction, as in the $\phi^4$ model with spontaneous symmetry breaking, and nonpolynomial interaction, as in the sine-Gordon model.

A particularly interesting result of this work concerns scenarios where the brane may split, engendering internal structure. The mechanism used to describe the brane splitting is different from other descriptions, where thermal effects \cite{S1} and the presence of 2-kink solutions \cite{S2} play the role for the splitting, in scenarios with standard gravity.

%%%%%%%%%%%%%%%%%%%%%%%%%%%%
\section{Generalities}

We start with a $5D$ action which describes a generalized $F(R)$ brane, in which gravity is coupled to a real scalar field $\phi$ in the form
\begin{equation}\label{eq1}
S\!=\!\int d^5x\sqrt{|g|\,}\left(-\frac14F(R)+{\cal L}(\phi,\nabla_a\phi)\right),
\end{equation}
where ${\cal L}(\phi,\nabla_{a}\phi)$ is the Lagrange density that accounts for the scalar field. It has the form
\begin{equation}\label{eq2}
{\cal L}\!=\!\frac{1}{2}g_{ab}\nabla^a\phi\nabla^b\phi-V(\phi).
\end{equation}
Here we are using $4\pi G^{(5)}\!=\!1$, $g\!=\!det(g_{ab} )$, and the signature of the metric is $(+----)$. Also, $V(\phi)$ is the potential, to be defined below. We take the spacetime coordinates and fields as dimensionless quantities, and we use Latin indices for the bulk coordinates, $a,b\!=\!0,1,2,3,4,$ and Greek indices for the embedded $(3+1)$-dimensional space, $\mu,\nu\!=\!0,1,2,3$.

We study the case of a flat brane, with the line element
\begin{equation}\label{eq3}
ds^2\!=\!e^{2A}\eta_{\mu\nu}dx^\mu dx^\nu-dy^2,
\end{equation}
where $e^{2A}$ is the warp factor, $\eta_{\mu\nu}$ is the 4-dimensional Minkowski metric, and $y\!=\!x^4$ is the extra dimension.  

As usual, in the braneworld framework we suppose that both $A$ and $\phi$ are static and depend only on the extra dimension, that is, we
set $A \!=\! A(y)$ and $\phi \!=\! \phi(y)$. In this case, the equation of motion for the scalar field has the form
\begin{equation}
\phi^{\prime\prime}+4A^{\prime}\phi^{\prime}\!=\!V_{\phi},
\end{equation}
where prime denotes derivative with respect to the extra dimension, and $V_{\phi}\!=\!dV/d\phi$.

The energy density $\rho$, that is, the $T_{00}$ component of the energy-momentum tensor is given by
\begin{equation}
\rho\!=\!-e^{2A(y)}{\cal L}.
\end{equation}
For static fields the modified Einstein equations becomes
\bes\label{eq4}
\ben
&&\!\!\!\!\!\!  -\frac23 \phi^{\prime 2}\!=\!A^{\prime\prime} F_R- \frac13 A^\prime F_{R}^\prime + \frac13 F_{R}^{\prime\prime} , \label{eq4.1}\\
&&\!\!\!\!\!\! V(\phi)\!-\!\frac12 \phi^{\prime2}\!=\!2(A^{\prime\prime}\!+\!A^{\prime2}) F_R \!-\! \frac14 F(R)\!-\!2A^\prime F_{R}^\prime,
 \label{eq4.2}
\een
\ees 
where $F_R\!=\!dF/dR$, $F_{RR}\!=\!d^2F/dR^2$, etc.

It was shown in \cite{BMP} that if the scalar curvature is constant (so, $R^{\prime}\!=\!0$), it  is possible to obtain analytic solutions for the field equations, in the case of several scalar fields. Here, however, we shall study more general solutions, that is, we shall assume that the scalar curvature is a generic function of the extra dimension, $R\!=\!R(y)$. Therefore, from  Eq. \eqref{eq4.1} we have
\bes
\begin{equation}\label{eq5}
\phi^{\prime2}\!=\! -\frac32 A^{\prime\prime}F_R\!+\! \frac12 \left(A^\prime R^\prime \!-\! R^{\prime\prime}\right)F_{RR}\!-\!\frac12 F_{RRR} R^{\prime2}.
\end{equation}
Also, the potential can be found from \eqref{eq4.2} and reads
\ben
 V(\phi)\!\!\!\!\!\!&&=- \frac14 F(R)+\frac14\left(5A^{\prime\prime}+8 A^{\prime2}\right) F_R -\nn
 &&- \frac14 (7A^\prime R^\prime+R^{\prime\prime})F_{RR}-\frac14 F_{RRR}R^{\prime2}.\label{eq6}
\een
\ees 
The scalar curvature is given in terms of the warp factor as
\be \label{eq7}
R\!=\!8A^{\prime\prime}+20A^{\prime 2}.
\ee
This can be used to rewrite Eqs.~\eqref{eq5} and \eqref{eq6} as
\bes \label{eq8}
\ben
\phi^{\prime2}
\!\!\!&=&\!\!\!-\frac32 A^{\prime\prime}F_R\!+\! 4\left(5A^{\prime2}A^{\prime\prime}\!  -\! 5 A^{\prime\prime2} \!-\! 4 A^\prime A^{\prime\prime\prime}\!-\! A^{\prime\prime\prime\prime}\right)\!F_{RR}\!\nn
&&-32F_{RRR} \left(5A^{\prime}A^{\prime\prime}\! + \!A^{\prime\prime\prime}\right)^2 , \label{eq8.1}\\
 V(\phi)\!\!&=&\!\!- \frac14 F(R)+ \frac14\left(5 A^{\prime\prime}  +8 A^{\prime2}\right) F_R-(70A^{\prime2}A^{\prime\prime} \!+\!\nn
&&\left. + 24A^{\prime\prime\prime}A^\prime  \!+ \!10 A^{\prime\prime2} \!+\! 2 A^{\prime\prime\prime\prime}\right)F_{RR} -\nn
& &- 16 F_{RRR} \left(5A^{\prime}A^{\prime\prime} + A^{\prime\prime\prime}\right)^2.\label{eq8.2}
\een
\ees 
It is not hard to check that in the simplest case where $F(R)=R$, these equations reduce to the standard result; see, e.g., \cite{B}. However, in the general case these equations depend on
the third and fourth derivative of the warp function, that is, on $A^{\prime\prime\prime}$ and $A^{\prime\prime\prime\prime}$.}

To find explicit solutions, we follow \cite{AB} and choose the simplest nontrivial polynomial function $F(R)\!=\!R+\alpha R^2$,  where $\alpha$ is a real parameter.
In this case we can write \eqref{eq8} as 
\bes \label{eq9}
\ben
&&\!\!\!\!\phi^{\prime2}
\!=\! -\frac32 A^{\prime\prime}\!\!-\!4\alpha \!\left(5A^{\prime\prime}\!A^{\prime2}\!\!+\!16A^{\prime\prime2}\!\!+\!8 A^\prime A^{\prime\prime\prime}\!\!+\!2A^{\prime\prime\prime\prime}\right),\label{eq9.1}\\
&&\!\!\!\!V(\phi)\!=\! -3A^{\prime2} \!-\!\frac34 A^{\prime\prime}\!-\!2\alpha \left(10A^{\prime4}\!+\!69A^{\prime2}A^{\prime\prime}\!+\!24A^{\prime}A^{\prime\prime\prime}\!+\right.\nn
&&\;\;\;\;\;\;\;\;\;+\left.8A^{\prime\prime2}\!+\!2A^{\prime\prime\prime\prime}\right).\label{eq9.2}
\een
\ees
Also, the energy density can be written in terms of the warp function as
\ben \label{eq10}
\rho&=&-e^{2A}\left[\frac32 A^{\prime\prime}+3A^{\prime2}+4 \alpha\left(5A^{\prime4}\!+\!37A^{\prime2}A^{\prime\prime}\!+\right.\right.\nn
&&\left.\left.+~16A^{\prime}A^{\prime\prime\prime}\!+12A^{\prime\prime2}\!+
\!2A^{\prime\prime\prime\prime}\right)\right],
\een
or
\ben \label{eq10.1}
\rho\!\!\!&=&\!\!\!-\frac{d}{dy}\!\left[\!e^{2A}\!\left(\frac32A^{\prime}\!+\!4\alpha\left(\!\frac{13}3 A^{\prime 3}\!+\!2 A^{\prime\prime\prime}\!+\!12A^{\prime}A^{\prime\prime}\!\right)\!\!\right)\!\right]+\nn
&&+~\frac{44\alpha }{3} e^{2A}A^{\prime 4}.
\een
If $\alpha\!=\!0$, the energy density becomes $\rho=-(3/2)(e^{2A} A^\prime)^\prime$.
%\ben \label{eq10.2}
%\rho\!\!&=&\!\!-\frac32\frac{d}{dy}\left(e^{2A}A^{\prime}\right).
%\een
It is a total derivative, so the energy vanishes in the standard situation. If $\alpha \neq 0$, the first term in equation \eqref{eq10.1} do not contribute to the energy, so we can write
\ben\label{eq10.3}
E\!=\!\frac{44}{3}\,\alpha  \int dy ~e^{2A}A^{\prime 4},
\een
and the sign of $\alpha$ controls the sign of the energy.

%%%%%%%%%%%%%%%%%%%%%%%%%%%%
\section{Exact Procedure}

\begin{figure}[!ht] 
\includegraphics[scale=0.58]{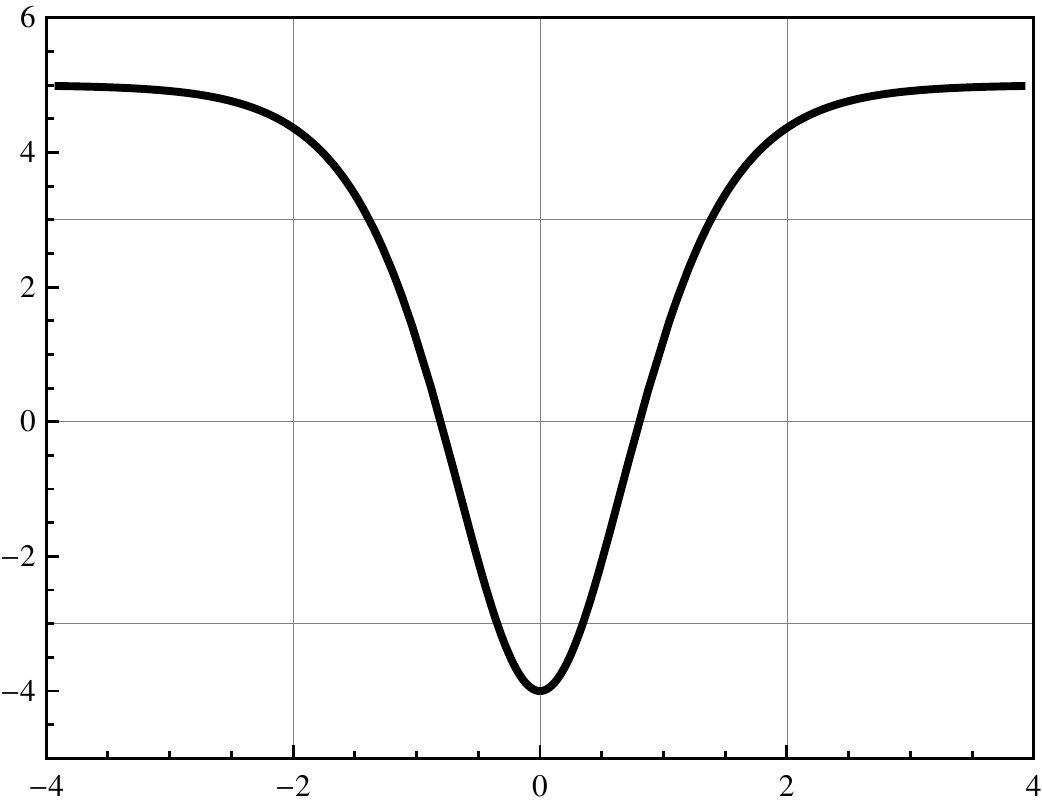}
\vspace{-0.4cm}
\caption{\small{ The scalar curvature, depicted from Eq.~\eqref{eq12} for $k=1$ and $B=1/2$}.}
\label{figure1}
\end{figure}

Let us start studying a model that engenders exact results. To do this, we choose the following warp function \cite{G}
\begin{equation}\label{eq11}
A(y)\!=\!B\ln \left[\sech(k y)\right],
\end{equation}
where $B$ and $k$ are positive parameters. This allows us to write $R$ as
\begin{equation}\label{eq12}
R(y)\!=\!4Bk^2\Big[5B-(5B+2)\sech^2(ky)\Big].
\end{equation}
The scalar curvature \eqref{eq12} is depicted in Fig. \ref{figure1}. In the limit $y\to\pm\infty$, $R$ tends to a constant value, $20 B^2 k^2$, and at $y=0$, the scalar curvature becomes $-8Bk^2$.

%%%%%%%%%%%%%%%%%%%%%%%%%%%%
%%%%%%%%%%%%%%%%%%%%%%%%%%%%

For the warp function \eqref{eq11}, the expressions \eqref{eq9} reduce to
\bes \label{eq13}
\ben
\!\frac{\phi^{\prime2}}{Bk^2}\!\!\!\!&=&\!\!\!\!\frac32 S_k^2\!\!+\!4\alpha k^2 \!S_k^2\!\left[5B^2\!\!+\!16 B\!+\!8\!-\!(5B^2\!\!+\!32B\!+\!12)\!S_k^2\right]\!\!, \nn\label{eq13.1}\\
\!\frac{V(y)}{Bk^2}\!\!\!\!&=&\!\!\!\!-3B\!+\!3\!\left(\!\!B\!+\!\frac14\!\right)S_k^2\!-\!2\alpha k^2 \!\!\left[\!10 B^3\!-\!(20B^3\!+\!69B^2\!+\right.\nn
&&\!\!\!\!+\!\left.\!48B\!+\!8)S_k^2\!\!+\!\left(\!10B^3\!\!+\!69B^2\!\!+\!56B\!+\!12\right)\!S_k^4 \right]\!.\label{eq13.2}
\een
\ees 
Here we are using $S_k\!=\!S(ky\!)=\!\sech(ky)$, for simplicity. As one knows, solutions $\phi=\phi(y)$ of the above equation \eqref{eq13.1} must go to some constant value $\bar\phi$ asymptotically. Thus, if we want that our model makes physical sense, the potential \eqref{eq13.2} should go to a vacuum value for $\bar\phi$. In the general case, the asymptotic values of the potential and its derivative with respect to the field are, respectively,
\begin{equation}\label{potcond}
\Lambda_5\equiv V\!(\phi\!\!\to \!\!\pm \bar\phi)= \!\!-B^2k^2(3+\!20\alpha B^2k^2)
\end{equation}
and $V_{\phi}(\phi\!\!\to \!\!\pm \bar\phi)\!\!\rightarrow\! 0$. 

If we use Eq.~\eqref{eq13.1}, we can infer the range of values for the parameter $\alpha$: since $\phi$ and its derivative are real, one must have $\phi^{\prime2}\!>\!0$ and this implies the following condition
\begin{equation}\label{eq15}
\frac{-3}{8k^2(8\!+\!16B\!+\!5B^2)}\!=\!\alpha_1\!\leq\!\alpha\!\leq\!\alpha_2\!=\!\frac{3}{32k^2(1\!+\!4B)}.
\end{equation}
Furthermore, the energy density can be written as
\ben\label{eq16}
\frac{\rho}{Bk^2} \!\!&=&\!-3BS_k^{2B}\!+\!3\!\left(\!B\!+\!\frac12\!\right)S_k^{2B+2}\!-\nn
&&\!-4\alpha k^2 S_k^{2B}\Bigg[5B^3\!-\!\left(10 B^3\!+\!37B^2\!+\!32B\!+\!8\right)S_k^2\!+\nn
&&+\!\left(5 B^3\!+\!37B^2\!+\!44B+12\right)S_k^{4}\Bigg],
\een 
and for $\alpha\!=\!0$ it becomes
\ben
\rho(y) \!=\!-3B^2k^2S_k^{2B}+3Bk^2\left(B\!+\!\frac12\right)S_k^{2B+2}.
\een 

We see that the energy density \eqref{eq16} displays the presence of an inflection point at $y\!=\!0$, such that
\begin{equation}\label{eq18}
\alpha_s\!=\!\frac{3+9B}{8k^2(16+60B + 49 B^2)}.
\end{equation}
This means that at $\alpha\!=\!\alpha_s$ the solution of Eq.~\eqref{eq13.1} begins to split, inducing internal structure to the brane. Moreover, the energy density behaves asymptotically as
\begin{equation}
\rho(y)\!=\! -2^{2B}B^2k^2\Bigg(3+20\alpha k^2B^2 \Bigg)e^{-2kBy}+\cdots.Moreover
\end{equation}

\begin{figure}[!ht]
\begin{center}
\includegraphics[scale=0.6]{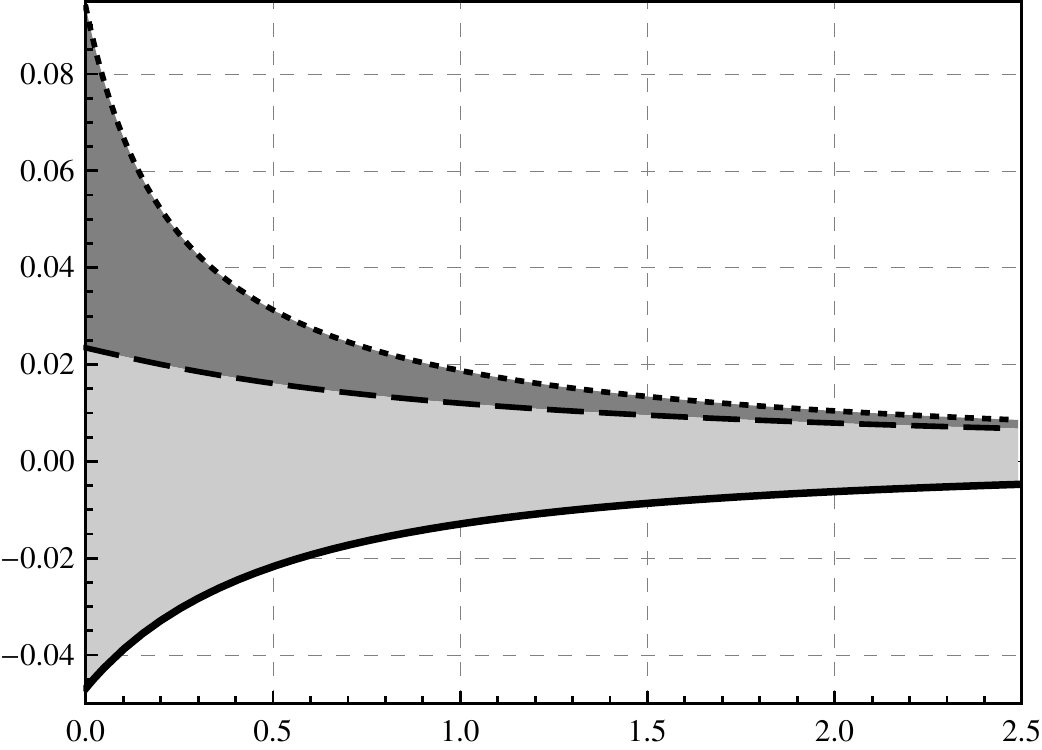}
\includegraphics[scale=0.58]{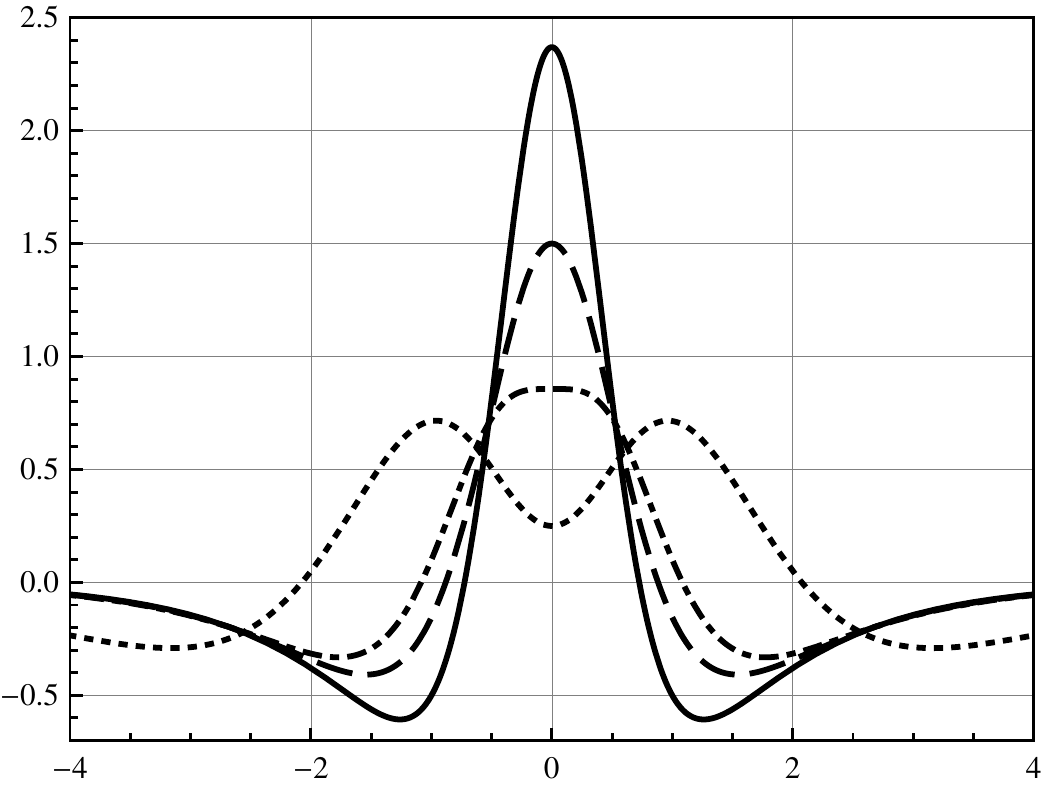}
\end{center}
\caption{\small{ Upper panel: the region bounded by $\alpha$ in Eq.~\eqref{eq15}, for $k=1$ and $\alpha=-1/46$ (solid line), $\alpha=15/932$ (dashed line) and $\alpha=1/32$ (dotted line). Lower panel: the energy density \eqref{eq16} obtained for $k=1$ and $B=1/2$, and for $\alpha=-1/46$ (solid line), $\alpha=0$ (dashed line), $\alpha=15/932$ (dotted-dashed line) and $\alpha=1/32$ (dotted line).}}\label{figu2}
\end{figure}

In Fig.~$\ref{figu2}$, in the upper panel we depict the allowed region of $\alpha$ for distinct values of $B$ and $k\!=\!1$. The two gray regions follow in accordance with Eq.~\eqref{eq15}. The solid line is obtained for $\alpha\!=\!\alpha_1$, the dashed line for $\alpha\!=\!\alpha_s$ and the dotted line for $\alpha\!=\!\alpha_2$. In the light gray region the energy density has a maximum at $y\!=\!0$ and for $\alpha$ in the darker region the energy density has a local minimum at $y\!=\!0$. The dashed line represents the appearance of the inflection point, where the brane starts to split. The darker gray region identifies the region where the brane splits, engendering internal structure. To further illustrate this situation, in the lower panel in Fig.~$\ref{figu2}$ we depict the energy density \eqref {eq16} for $\alpha=-1/46, \alpha=0, \alpha=15/932$ and $\alpha=1/32$. The case $\alpha=1/32$ nicely illustrates the brane splitting.

%%%%%%%%%%%%%%%%%%%%%%%%%%%%
%%%%%%%%%%%%%%%%%%%%%%%%%%%%

The energy of the brane, given by Eq.~\eqref{eq10.3}, is
\ben\label{eq19}
E=11\alpha k^3\sqrt{\pi} \frac{B^4\Gamma(B)}{\Gamma(5/2 + B)}.
\een

We can solve Eq.~\eqref{eq13.1}, to find a function $\phi\!=\!f(y)$ that may be inverted to give $y\!=\!f^{-1}(\phi)$, which allows us to write the potential in the usual way $V\!=\!V(\phi)$. The general solution of the equation \eqref{eq13.1} looks like
\ben
\phi(y)&=&\frac{\sqrt{6B}}{2}\sqrt{1-\frac{\alpha}{\alpha_1}}~ F(ky, P)\,+\nn
&+&\!\!\!\!\frac{\sqrt{3B}}2\!\!\left[\!1\!-\!\frac{2\alpha_1\!-\!\alpha_2}{\alpha_1\alpha_2}\alpha \!+\!\frac{\alpha _1\!-\!\alpha }{\alpha _1}\!\cosh(2ky)\!\right]^{1/2}\! \!\!\!\!\!\!\tanh(ky)\nn
\een
where 
\be
F(y, P)=-\int^{y}_0\left[\frac{\sinh^2(x)}{ \sqrt{P+\sinh^2(x)}}\right]dx,
\ee
and
\be
P  =\frac{\alpha  \alpha _1-\alpha _1 \alpha _2}{\alpha  \alpha _2-\alpha _1 \alpha _2}.
\ee
The general result leads to some interesting particular cases. The first is the limiting case for $\alpha=\alpha_1$ in Eq.~\eqref{eq15}; in this case $P\gg 1$, and so we can write the solution $\phi$ and the potential as, respectively
\bes
\ben
\phi(y)\!\!\!&=&\!\!\! \sqrt{\frac{3B (6+B) (2+5 B)}{16+2B (16+5 B)}}\tanh(ky), \label{sola}\\
V(\phi)\!\!\!&=&\!\!\!\frac{3Bk^2}{4} \! \left(\!\frac{12\!+\!24 B+5B^2}{8\!+\!16B\!+\!5 B^2}\!\right)\!-\!k^2(1 + 2 B)  \phi^2+\nn
&&\!\!\!+~k^2\frac {(1 + 2 B) \left(8+16B+5 B^2\right)} {3 B (6 + B) (2 + 5 B)}\phi^4.\label{pota}
\een
\ees 

The second case is the standard case, where $\alpha\!=\!0$ and $P\!=\!1$; here we obtain
\bes
\ben
\phi(y)\!\!\!&=&\!\!\!\sqrt{6B} \arctan\Bigg[\tanh\Bigg(\frac{k y}{2}\Bigg)\Bigg], \label{solb}\\
V(\phi)\!\!\!&=&\!\!\!-3B^2k^2\!+\!3Bk^2\!\left(\!\!B\!+\!\frac14\!\right)\! \cos^2\!\left(\!\!\sqrt{\frac2{3B}}\phi\!\right)\label{potb}\!,
\een
\ees 
which reproduces the result obtained \cite{G}.

The third case concerns the value $\alpha=\alpha_2$. here we have $P=0$. Also,
\bes
\ben
\!\!\!\!\!\!\!\!\!\phi(y)\!\!\!&=&\!\!\!\sqrt{\frac{3B (6\!+\!B) (2\!+\!5 B)}{8\!+\!32B}}~\!\left[1\!-\!\sech(ky)\right]\!{\rm sign }(y) \label{solc}, \\
\!\!\!\!\!\!\!\!\!V(\phi)\!\!\!&=&\!\!\!-\frac{3B^2k^2}{8}\!\left(\!\frac{8\!+\!32 B\!+\!5 B^2}{1\!+\!4B}\!\right)\!+\!\frac{3Bk^2}{16} (6\!+\!B) (2\!+\!5 B)\times\nn
&&\!\!\!\times\left[1\!-\!\left(\frac{3B (6\!+\!B) (2\!+\!5 B)}{8+32B}\right)^{\!\!\!-1/2}\!\!\!\!\!|\phi|\right]^2-\nn
&&\!\!\!-\frac{3 Bk^2}{16}\frac{(6\!+\!B) (1\!+\!2B) (2\!+\!5B) }{1+4 B} \times\nn
&&\!\!\!\times\left[1\!-\!\left(\frac{3B (6\!+\!B) (2\!+\!5 B)}{8+32B}\right)^{\!\!\!-1/2}\!\!\!\!\!|\phi|\right]^4.\label{potc}
\een
\ees 

Note that each one of the particular solutions obtained above satisfies the following conditions: $\phi(y\!\!\to\!\! \pm \infty)\!\to\!\pm \bar{\phi}$, where
\ben
\bar{\phi}_1&=&\sqrt{\frac{3B (6+B) (2+5 B)}{16+2B (16+5 B)}}\nn
\bar{\phi}_0&=&\frac{\pi}{4}\sqrt{6B}\nn
\bar{\phi}_2&=&  \sqrt{\frac{3B (6+B) (2+5 B)}{8+32 B}}\nonumber
\een
with $\bar{\phi}_1<\bar{\phi}_0<\bar{\phi}_2$, for $\alpha\!=\!\alpha_1$, $\alpha\!=\!0$ and $\alpha\!=\!\alpha_2$, respectively.  Furthermore, for each one of the three cases we have $V(\bar{\phi})\!=\!v$ e $V_\phi(\bar{\phi})\!=\!0$, where
\ben
v_1&=&-3B^2k^2\left(\frac{16+32 B+5 B^2}{16+32 B+10 B^2}\right)\nn
v_0&=&-3 B^2 k^2\nn
v_2&=&-3B^2k^2\left(\frac{8+32 B+5 B^2}{8+32B}\right)\nonumber
\een
where $v_1<v_0<v_2$, which agrees with the previous result obtained in Eq.~\eqref{potcond}.

\begin{figure}[!ht]
\begin{center}
\includegraphics[scale=0.58]{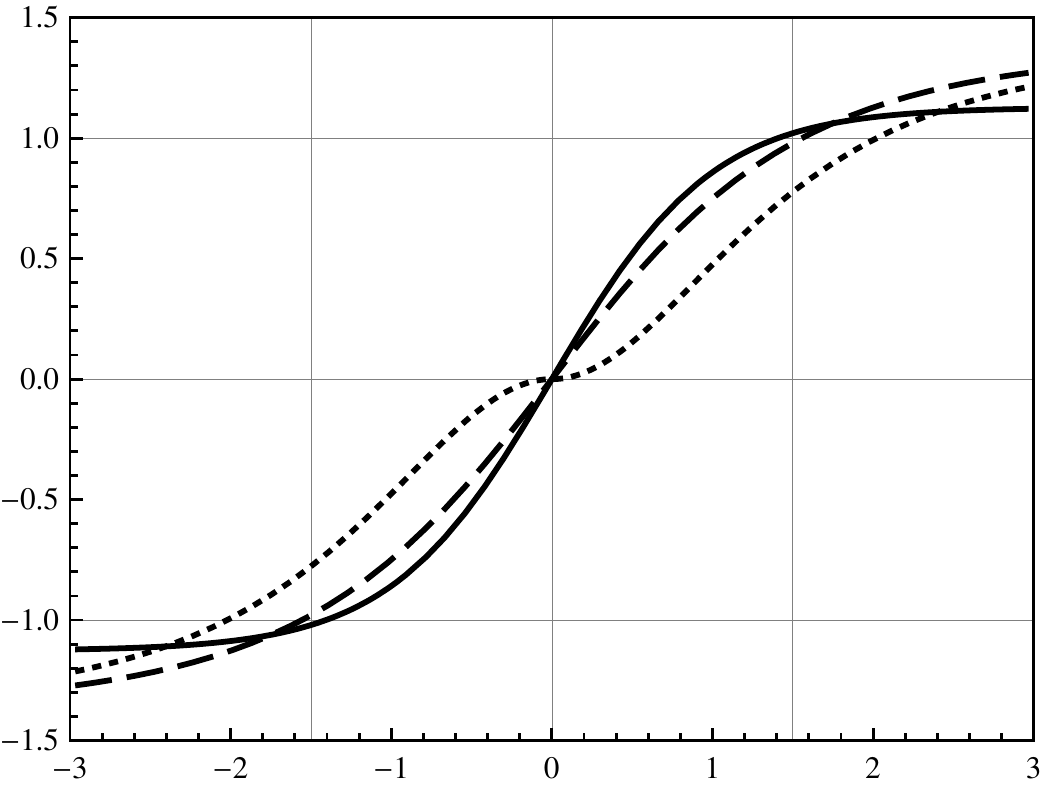}
\includegraphics[scale=0.58]{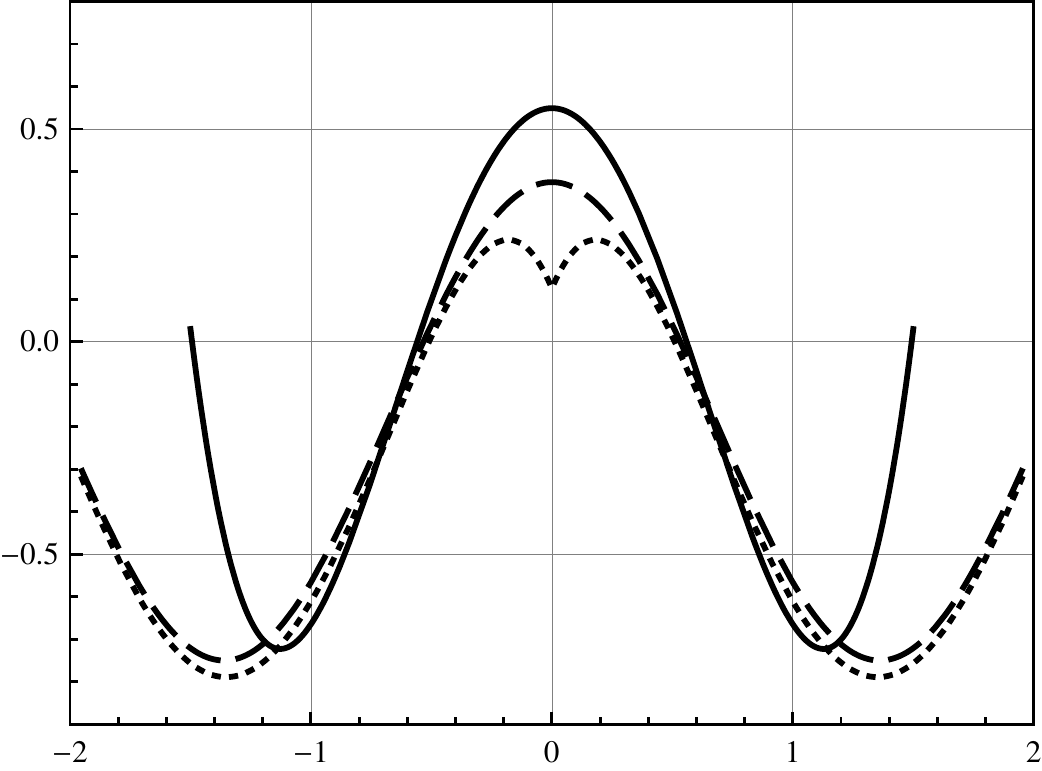}
\end{center}
\caption{\small{Upper panel: the three specific solutions \eqref{sola}, \eqref{solb} and \eqref{solc} for $k=1$, $B=1/2$ and $\alpha=-1/46$ (solid line), $\alpha=0$ (dashed line) and $\alpha=1/32$ (dotted line). Lower panel: the three specific potentials \eqref{pota}, \eqref{potb} and \eqref{potc}, for the same values of the parameters $k$, $B$ and $\alpha$.}}\label{figu3}
\end{figure}

In the upper panel in Fig.~\ref{figu3} we depict the three cases where solutions are obtained exactly. We depict the solutions for $k\!=\!1$ and $B\!=\!1/2$, for $\alpha\!=\!-1/46$ (line solid), $\alpha\!=\!0$ (dashed line) and $\alpha\!=\!1/32$ (dotted line). We see that the defect modifies, starting to behave as a 2-kink when $\alpha>\alpha_s(=\!15/932)$. In the lower panel in Fig.~\ref{figu3} we depict the potential, for the same values of the parameters $k$, $B$ and $\alpha$.

%%%%%%%%%%%%%%%%%%%%%%%%%%%%%%%%%%%%%%%%%%%%%%
%%%%%%%%%%%%%%%%%%%%%%%%%%%%%%%%%%%%%%%%%%%%%%

%%%%%%%%%%%%%%%%%%%%%%%%%%%%
\section{Perturbative Procedure}

In this section we follow the perturbative procedure introduced in \cite{ABLM} to investigate specific models. For this, we suppose that the field is expanded in the form $\phi(y)=\phi_0(y)+\alpha \phi_\alpha(y)$, Furthermore, the potential and the warp function are also written as $V(\phi)=V_0(\phi)+\alpha V_\alpha(\phi)$ and $A(y)=A_0(y)+\alpha A_\alpha(y)$; $\alpha$ is now a small parameter and $\phi_0$ and $A_0$ are the solution for $\alpha=0$, such that 
\begin{equation}\label{eq20}
\phi^{\prime}_0\!=\!\frac12 W_{\phi_0}, \,\,\,\, {\rm and}\,\,\,\,   A_0^{\prime}\!=\!-\frac13 W(\phi_0),
\end{equation}
where $W_{\phi_0}=W_{\phi}|_{\phi=\phi_0}$, etc. Also, 
\begin{equation}
V_0(\phi)\!=\!\frac18W_{\phi}^2\!-\!\frac13W^2.
\end{equation}

Using Eqs.~\eqref{eq9.1} and \eqref{eq9.2} we can write, up to first order in $\alpha$,
\bes
\ben
\!\!\!\!\phi_\alpha^{\prime}(\phi_0)\!\!&=&\!\!-\frac{3}2\frac{A_\alpha^{\prime\prime}}{W_{\phi_0}} \!+\!\frac43W_{\phi_0}\bigg(\frac5{18} W^2\!-\! \frac43 W_{\phi_0}^2\!-\nn
&-&\!\!\frac43W W_{\phi_0\phi_0}\!+\! W^2_{\phi_0\phi_0}\!+\!\frac{1}{2}W_{\phi_0}W_{\phi_0\phi_0\phi_0}\!\bigg)
\een
and
\ben
V_\alpha(\phi_0) \!\!&=&\!\!- W_{\phi_0} \left(\frac14 W_{\phi_0\phi_0}\!-\!\frac23 W\right)
\phi_\alpha\!+ \!2W A_\alpha^\prime\!-\!\frac34 A^{\prime\prime}_\alpha\!-\nn
&&\!\!-\frac{20}{81} W^4\!+\!\frac{23}{9}  W^2 W^2_{\phi_0}\!-\frac{8}{3} W W^2_{\phi_0}W_{\phi_0\phi_0}-\nn
&&\!\!-\frac{4}{9} W^4_{\phi_0}\!+\frac{2}{3}W_{\phi_0}^2W^2_{\phi_0\phi_0}+\frac1{3} W^3_{\phi_0}W_{\phi_0\phi_0\phi_0}.
\een
\ees

We assume that $A_\alpha^{\prime}$ is a function of the field $\phi_0$, with the specific form 
\begin{equation}\label{eq22a}
A_\alpha^\prime=\beta W(\phi_0)+\gamma W^3(\phi_0),
\end{equation}
where $\beta$ and $\gamma$ are real parameters. This allows us to write the previous relations as
\bes\label{eq21}
\ben
\phi_\alpha^{\prime}\!\!&=&\!\!\frac43W_{\phi_0}\bigg[-\frac{9}{16}\beta +\left(\frac5{18}-\frac{27}{16}\gamma \right) W^2\!-\! \frac43 W_{\phi_0}^2\! -\nn
&&\!\!\!\!-\frac43W W_{\phi_0\phi_0}\!+\! W^2_{\phi_0\phi_0}+\frac{1}{2}W_{\phi_0}W_{\phi_0\phi_0\phi_0}\bigg],\label{eq21.1}
\een
and
\ben
V_\alpha(\phi_0) \!\!\!\!&=&\!\!\!\!- W_{\phi_0}\!\! \left(\frac14 W_{\phi_0\phi_0}\!-\!\frac23 W\right)
\phi_\alpha\!+ \! 2\beta W^2\!-\!\frac{3\beta}8 W_{\phi_0}^2\!+\nn
&&\!\!\!\!+2\bigg(\gamma\!-\frac{10}{81}\bigg) W^4\!+\bigg(\frac{23}{9}-\frac{9\gamma}8\bigg)  W^2 W^2_{\phi_0}-\nn
&&\!\!\!\!-\frac{8}{3} W W^2_{\phi_0}W_{\phi_0\phi_0}-\frac{4}{9} W^4_{\phi_0}\!+\frac{2}{3}W_{\phi_0}^2W^2_{\phi_0\phi_0}+\nn
&&\!\!\!\!+\frac1{3} W^3_{\phi_0}W_{\phi_0\phi_0\phi_0}.\label{eq21.2}
\een
\ees
We can thus reconstruct the potential
\ben
V(\phi) \!\!\!\!&=&\!\!\frac18W_{\phi}^2\!-\!\frac13W^2+\alpha \bigg[W_{\phi}\!\! \left(\frac23 W\!-\!\frac14 W_{\phi\phi}\right)\phi_\alpha\!+\nn
&+& \! 2\beta W^2\!-\!\frac{3\beta}8 W_{\phi}^2\!+\!2\bigg(\gamma\!-\frac{10}{81}\bigg) W^4+\nn
&&\!\!\!\!\!+\bigg(\frac{23}{9}-\frac{9\gamma}8\bigg)  W^2 W^2_{\phi}\!-\!\frac{8}{3} W W^2_{\phi}W_{\phi\phi}-\nn
&&\!\!\!\!-\frac{4}{9} W^4_{\phi}\!+\frac{2}{3}W_{\phi}^2W^2_{\phi\phi}\!+\!\frac1{3} W^3_{\phi}W_{\phi\phi\phi}\bigg].\label{eq21.3}
\een

So we can solve Eq.~\eqref{eq21.1} to obtain $\phi_\alpha (\phi_0)$ and therefore the potential $V_\alpha(\phi_0)$ in \eqref{eq21.2}. Furthermore, for the choice \eqref{eq22a} the warp function is obtained as
\begin{equation}
A(\phi_0)\!=\!-\frac23(1\!-\!3\alpha\beta)\!\!\int\!\!\frac{W}{W_{\phi_0}}d\phi_0\!+\!2\alpha\gamma\!\!\int\!\!\frac{W^3}{W_{\phi_0}}d\phi_0,
\end{equation}
and the constant of integration is obtained from the condition $A(0)=0$. 

To see how the above procedure works, let us now illustrate the results with two very distinct examples, one with polynomial potential, and the other with nonpolynomial potential.

%%%%%%%%%%%%%%%%%%%%%%%%%%%%%%%%%
\subsection{Polynomial potential}

In this first example, we choose the following function
\be\label{eq22}
W({\phi})\!=\!2\phi-\frac23 \phi^3.
\ee
In the absence of gravity, this choice represents the well-known $\phi^4$ model, with spontaneous symmetry breaking. With this, we can write the warp function as
\ben\label{eq24}
A(\phi_0)\!\!\!&=&\!\!\!-\frac13\left(\frac13\!+\!\frac{32}{9}\alpha\gamma\!-\! 
 \alpha\beta \right)\phi_0^2\!+\! \frac{38}{27}\alpha\gamma \phi_0^4\!-\!\frac{32}{81}\alpha\gamma \phi_0^6\!+\nn 
 &&\!\!\! +\frac{1}{27}\alpha\gamma\phi_0^8\!+\!\frac23\left(\frac13 \!-\! 
 \frac{16}{9}\alpha\gamma \!-\! \alpha\beta\right) \ln\left(1\! -\! \phi_0^2\right),
\een
which obeys $A(0)\!=\!0$. For the scalar field, we get its first-order corrections as 
\bes\label{eq26}
\begin{eqnarray}
\phi_{\alpha}(\phi_0)\!\!&=&\!\!\!-\bigg(\!\frac{224}{9}\!+\!\frac{3\beta}{2}\!\bigg)\phi_0\!-\!\bigg(\!6\gamma\!-\!\frac{3056}{81}\!\bigg) \phi_0^3\!-\nn
&&\!\!\!-\!\bigg(\!\frac{416}{81}\!-\!\frac{12\gamma}{5}\!\bigg)\phi_0^5\!-\!\bigg(\!\frac{2\gamma}{7}\!-\!\frac{80}{1701}\!\bigg)\phi_0^7 \label{eq26.1}.
\end{eqnarray}
Moreover, the first-order corrections to the potential become
\ben
\!V_\alpha\!(\phi_0)\!\!\!&=&\!\!\!-\!\frac{160}{9}\!-\!\frac{3 \beta }{2}\!+\!\left(\frac{3056}{27}\!-\!18 \gamma \!+\!4 \beta \right) \phi_0^2\!+\nn
\!\!\!\!&+&\!\!\!\!\left(\!\!52 \gamma \!+\!\frac{3\beta}{2}\!-\!\frac{38336}{243}\!\right)\!\! \phi_0^4\!+\!\!\left(\!\!\frac{53600}{729}\!-\!\frac{632 \gamma}{15}\!-\!\frac{4 \beta}{9}\!\right) \!\!\phi_0^6\!+\nn
\!\!\!\!&+&\!\!\!\!\left(\frac{52\gamma}{3}\!-\!\frac{992}{81}\right) \!\phi_0^8\!+\!\left(\frac{4624}{15309}\!-\!\frac{2854 \gamma }{945}\right) \!\phi_0^{10}\!+\nn
\!\!\!\!&+&\!\!\!\!\left(\frac{80\gamma}{567}-\frac{320}{45927}\right)\phi_0^{12}.
\een
\ees 
and so the potential \eqref{eq21.3} becomes
\ben
\!\!\!\!\!V\!(\phi)\!\!\!&=&\!\!\!\frac{1}{2}\!-\!\frac{7 }{3}\phi^2\!+\!\frac{25 }{18}\phi^4\!-\!\frac{4}{27}\phi^6\!+\!\alpha\bigg[\!-\!\frac{160}{9}\!-\!\frac{3 \beta }{2}\!+\nn
&+&\!\!\!\left(\frac{3056}{27}\!-\!18 \gamma \!+\!4 \beta \right) \phi^2\!+\!\left(\!\!52 \gamma \!+\!\frac{3\beta}{2}\!-\!\frac{38336}{243}\!\right)\!\! \phi^4\!+\nn
\!\!\!\!&+&\!\!\!\left(\!\!\frac{53600}{729}\!-\!\frac{632 \gamma}{15}\!-\!\frac{4 \beta}{9}\!\right) \!\!\phi^6\!+\!\left(\frac{52\gamma}{3}\!-\!\frac{992}{81}\right) \!\phi^8\!+\nn
\!\!\!\!&+&\!\!\!\left(\frac{4624}{15309}\!-\!\frac{2854 \gamma }{945}\right) \!\phi_0^{10}\!+\!\left(\frac{80\gamma}{567}-\frac{320}{45927}\right)\phi^{12}\bigg].\nn\label{eq26.2}
\een

%%%%%%%%%%%%%%%%%%%%%%%%%%%%
\begin{figure}[!ht]
\begin{center}
\includegraphics[scale=0.56]{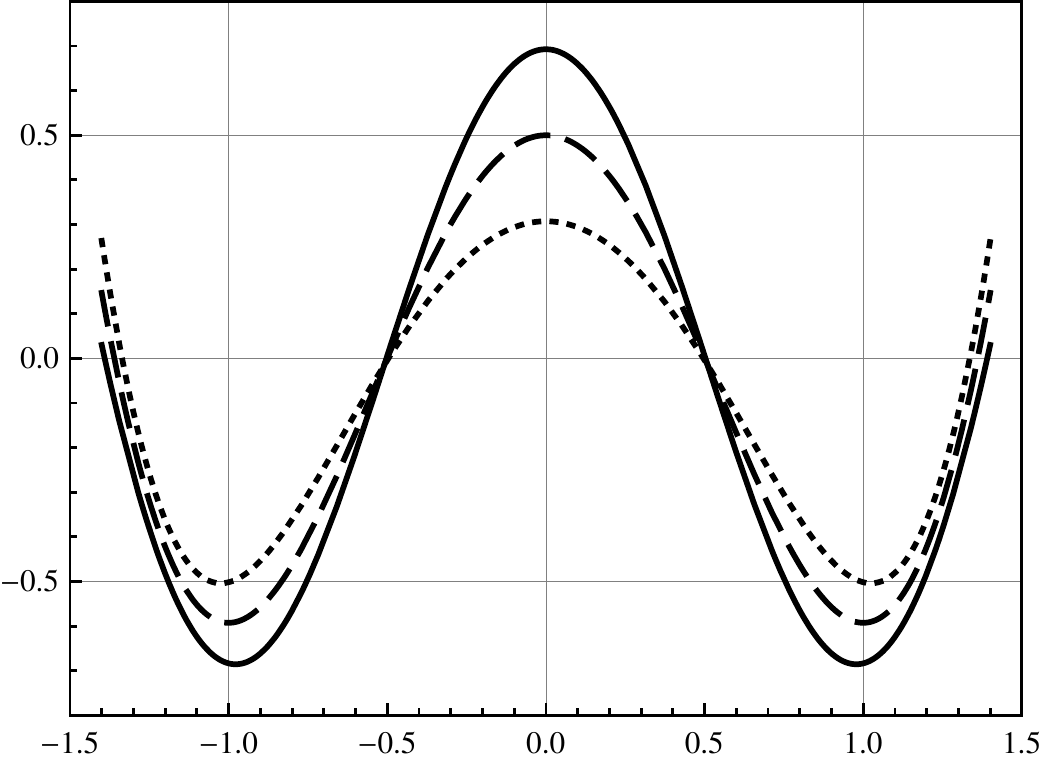}
\includegraphics[scale=0.56]{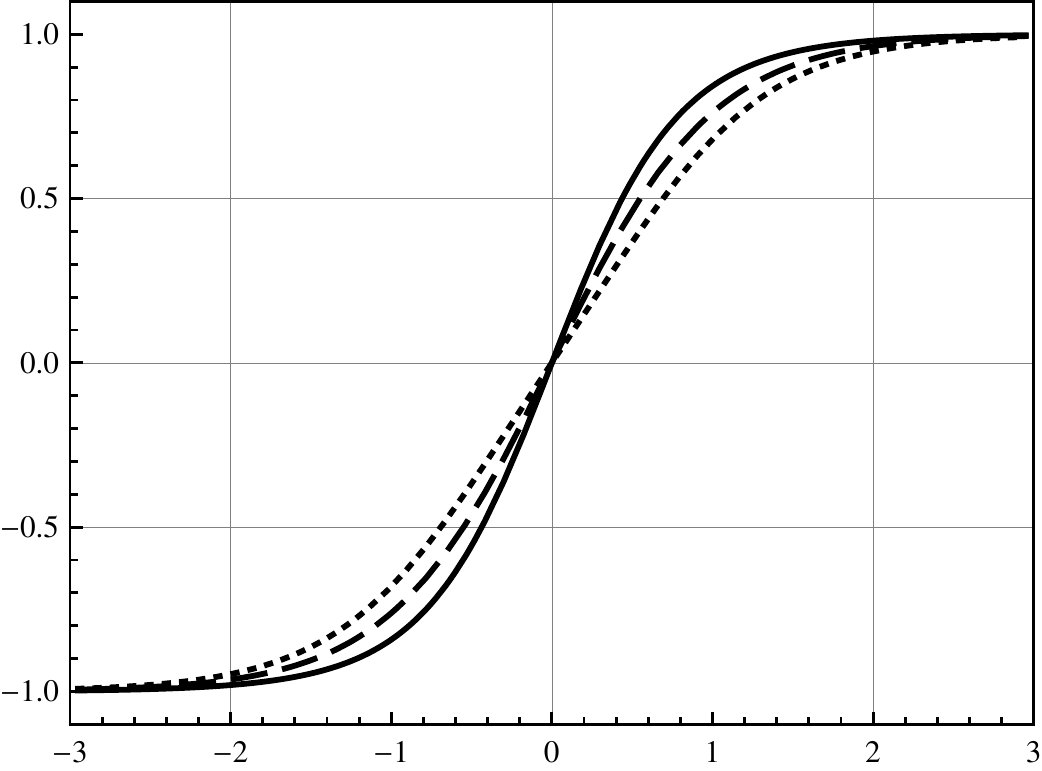}
\end{center}
\caption{\small{Upper panel: the potential \eqref{eq26.2} for $\gamma\!=\!1.6$, $\beta\!=\!1$ with $\alpha\!=\!-0.01$ (solid line), $\alpha\!=\!0$ (dashed line) and $\alpha\!=\!0.01$ (dotted line). Lower panel: the kinklike solution \eqref{eq27} for the same values of parameters.}}\label{figu4}
\end{figure}
%%%%%%%%%%%%%%%%%%%%%%%%%%%%

The model obtained with $\alpha\!=\!0$ has solution given by
\begin{equation}\label{sol1}
\phi_0(y)\!=\!\tanh(y),
\end{equation}

Using Eq.~\eqref{eq26.1}, we can write the complete solution $\phi(y)$ as
\ben
\phi(y)\!\!\!&=&\!\!\!\tanh(y)\!-\!\alpha\bigg[\bigg(\!\frac{224}{9}\!+\!\frac{3\beta}{2}\!\bigg)\!\tanh(y)\!+\nn
&&\!\!\!+\!\bigg(\!6\gamma\!-\!\frac{3056}{81}\!\bigg)\! \tanh^3(y)\!+\!\bigg(\!\frac{416}{81}\!-\!\frac{12\gamma}{5}\!\bigg)\!\tanh^5(y)\!+\nn
&&\!\!\!+\!\bigg(\!\frac{2\gamma}{7}\!-\!\frac{80}{1701}\!\bigg)\!\tanh^7(y) \bigg].\label{eq27}
\een
It obeys $\phi(y\!\rightarrow\! \pm \infty)\!\rightarrow \!\pm\bar\phi$ where
\be
\bar\phi\equiv\left[1\!+\!\alpha\left(\frac{13184}{1701}\!-\!\frac{136\gamma}{35}\!-\!\frac{3\beta}{2}\right)\right].
\ee
We see that $\phi$ is shifted asymptotically by some value, which depends on $\alpha$, as shown in the expression above. We also see that if $\beta\!\!=\!\!(16/25515)\times (8240\!-\!4131 \gamma )$ the field $\phi$ converges to the same asymptotic value of the field $\phi_0$, i.e., $\phi(y\!\!\to\!\!\pm\infty)\!=\!\phi_0(y\!\!\to\!\!\pm\infty)$.
In Fig.~\ref{figu4}, in the upper and lower panels, we depict the potential and kinklike solutions, respectively, for some small values of $\alpha$. 	
Furthermore, we have
\begin{equation}
V(\phi\!\rightarrow\!\!\bar\phi)=-\frac{16}{27}+ \frac{32\alpha}{6561} \left(1296\gamma+729\beta-160\right).
\end{equation}

Here we note that the minima of the potential are displaced up or down from the standard situation with $\alpha=0$ by the value
\be\nonumber
\triangle V_{\mbox{\small{min}}}= \frac{32\alpha}{6561} \left(1296\gamma+729\beta-160\right).
\ee
Similarly, the maxima are also displaced up or down by
\be\nonumber
\triangle V_{\mbox{\small{max}}}=-\alpha \left(\frac{160}{9}+\frac{3 \beta }{2}\right).
\ee
We can also note that $V_{\phi}(\phi\!\rightarrow\!\!\bar\phi)=0$.

\begin{figure}[!ht]
\begin{center}
\includegraphics[scale=0.56]{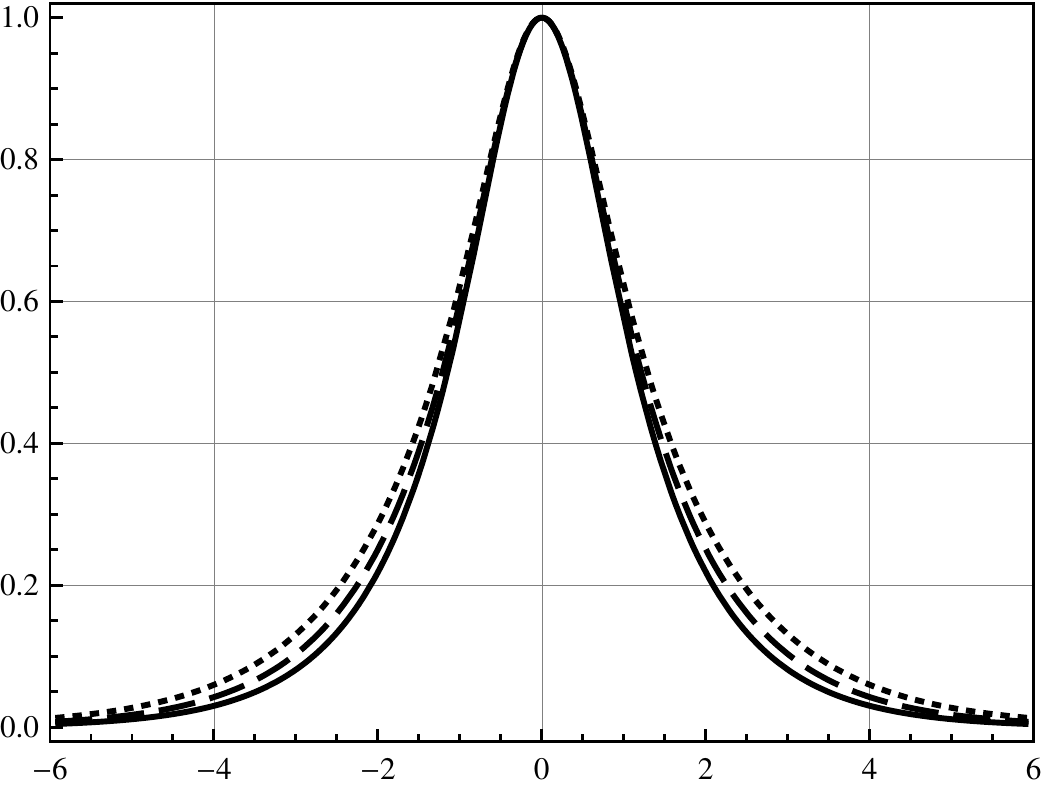}
\includegraphics[scale=0.56]{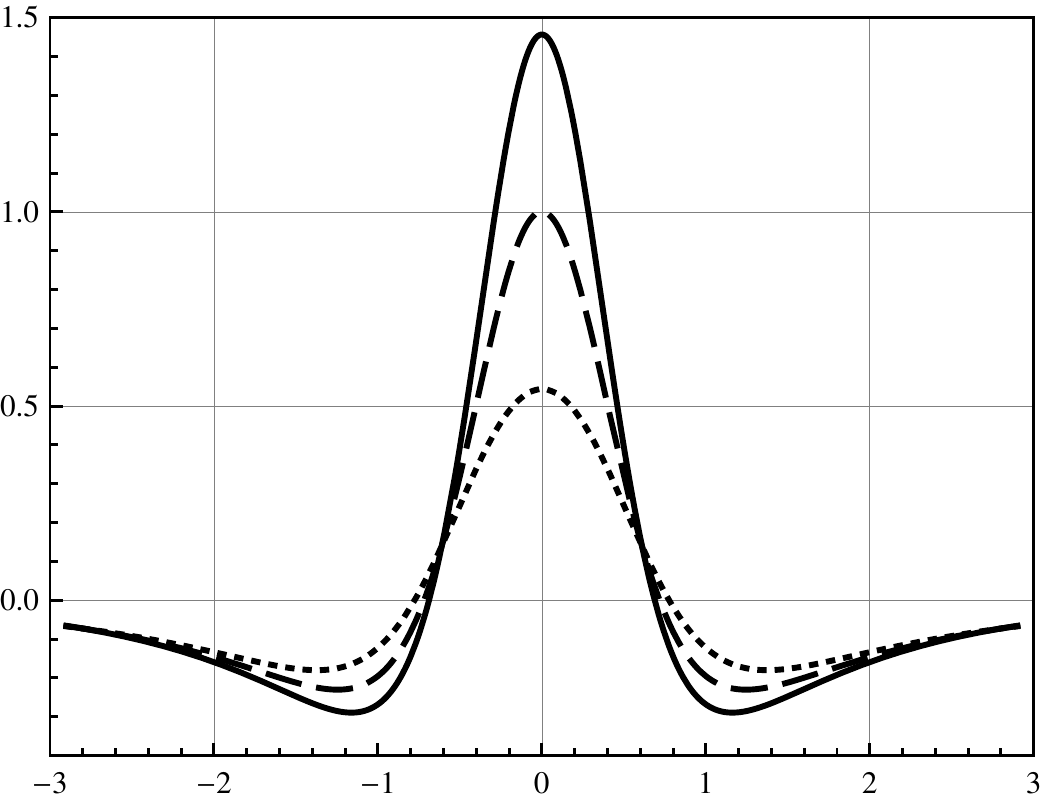}
\end{center}
\caption{\small{Upper panel: warp factor given by \eqref{eq28} for $\gamma\!=\!1.6$ and $\beta\!=\!1$ with $\alpha\!=\!-0.01$ (solid line), $\alpha\!=\!0$ (dashed line) and $\alpha\!=\!0.01$ (dotted line). Lower panel: energy density \eqref{eq30} for the same values of parameters.}}\label{figu5}
\end{figure}

With the solution \eqref{sol1} the warp function becomes
\ben
A(y)\!\!\!&=&\!\!\!\frac13\bigg(\!\frac{1}{3}\!-\!\frac{16}{9} \alpha  \gamma \!-\! \alpha  \beta \!\bigg)\bigg(S^2\!+\!4\ln S\bigg)\!+\nn
& &\!\!\!+\frac{4}{9}\alpha\gamma S^4\!+\!\frac{20}{81}\alpha\gamma S^6\!+\!\frac{\alpha\gamma}{27}S^8.\label{eq28}
\een
where $S\!=\!S(y)\!=\!\sech(y)$. This shows that, to make $e^{2A}>0$ we must impose that $\alpha$ is bigger (lesser) than $ 3(16 \gamma+9\beta)^{-1}$ for $\gamma$ lesser (bigger) than $ -9\beta/16$.
In Fig.~ \ref{figu5}, in the upper panel we depict the warp factor $e^{2A}$ for some values of the parameters, and in the lower panel we depict the energy density \eqref{eq30} for the same values of parameters.

%%%%%%%%%%%%%%%%%%%%%%%%%%%%
%%%%%%%%%%%%%%%%%%%%%%%%%%%%

Using the equations \eqref{eq26.2} and \eqref{eq27} we can find the energy density as
\ben\label{eq30}
\rho(y)\!\!\!&=&\!\!\!\frac{e^{\frac{2}{9} S^2}}{27 e^{\frac{2}{9}}}S^{8/9} \left(4S^6+39 S^4-16\right)+\nn
&&\!\!\!+\frac{\alpha S^{8/9}}{e^{\frac{2}{9}(1\!-\!S^2)}}\!\bigg[\!\sum_{i=0}^{7}\!a_{2i}S^{2i}\!\!+\!\ln(S)\bigg(\!b_2\!+\!\!\sum_{j=1}^{2}\!b_{2j+2}S^{2j+2}\!\bigg)\!\bigg].\nn
\een
where

\begin{align}
&a_0\!=\!\frac{14176\gamma}{2187}\!+\!\frac{256\beta}{81}\!-\!\frac{5120}{6561} &&a_{10}\!=\!\frac{3008}{5103}\!-\!\frac{128048\gamma}{25515} \nn
&a_2\!=\!\!\frac{448256 }{15309}\!-\!\frac{356624\gamma}{25515}\!-\!\frac{427}{81} &&a_{12}\!=\!\frac{4918\gamma}{15309}\!-\!\frac{320}{45927} \nn
&a_4\!=\!\frac{10432}{189}\!-\!\frac{188906\gamma}{25515}\!-\!\frac{10\beta}{27} &&a_{14}\!=\!\frac{8\gamma}{729} \nn
&a_6\!=\!-\!\frac{4144160}{45927}\!-\!\frac{142804\gamma}{76545}\!-\!\frac{34\beta}{81} &&b_{2}\!=\!\frac{128}{81}\bigg(\frac{16\gamma}{9}\!+\!\beta\bigg) \nn
&a_8\!=\!-\frac{543968}{15309}\!+\!\frac{58244\gamma}{5103}\!-\!\frac{8\beta }{81}&&-\!\frac{16}{39}b_{4}\!=\!4b_{6}\!=\! b_{2}\nonumber
\end{align}

Asymptotically, we have
\begin{equation}
\rho(y)=-\frac{2^{8/9}}{e^{\frac{2}{9}}}\bigg(\frac{16}{27}-\alpha [a_{0} + b_2(-y+\ln 2)] \bigg)e^{-8y/9}+\cdots
\end{equation}

Now, if we use Eq.~\eqref{eq7}, we find the scalar curvature in the form
\ben
R(y)\!\!\!&=&\!\!\!\frac{16}{81}\bigg(20-42 S^4-5S^6\bigg) -\frac{16}{243}\alpha\bigg[640\gamma+360\beta\!-\nn
&&\!\!\!-3(752+171 \beta \gamma)S^4-10(32\gamma+9 \beta)S^6+ \nn
&&\!\!\!+1332\gamma S^8+\!564\gamma S^{10}+40 \gamma S^{12}\bigg].
\een
We see that asymptotically, the scalar curvature goes to the constant value
\be
R\rightarrow \frac{320}{81}- \frac{640}{243}\alpha (16 \gamma+9 \beta).
\ee
At the origin it becomes
\begin{equation}
R(0)=-\frac{16}3+\frac{16}{81} \alpha  \left[752-90\beta +\gamma (171\beta -752)\right].
\end{equation}

%%%%%%%%%%%%%%%%%%%%%%%%%%%%%%%%%%%%%%%%%%%%%%
\subsection{Nonpolynomial potential}

Another relevant example is the sine-Gordon model, which is defined by the following function
\be \label{eq31}
W(\phi)=2 a\sin(b\phi)
\ee
where $a$ and $b$ are real parameters. The unperturbed solution is given by 
\begin{equation} \label{eq32}
\phi_0(y)=\frac1b\arcsin\left[\tanh(a b^2 y)\right].
\end{equation}

Using equations \eqref{eq21} we can write the first-order contributions to the solution and the potential as
\bes
\ben 
\phi_\alpha(\phi_0)\!\!&=&\!\!\!\bigg(\frac{40 a^2}{27}+\frac{8 a^2 b^4}{3}-9 a^2 \gamma -\frac{3 \beta }{2}\bigg)\phi_0-\nn
&&\!\!\!-\frac{a^2}{b}\bigg(\!\frac{20}{27}\!+\!\frac{64 b^2}{9}+4 b^4\!-\!\frac{9 \gamma }{2}\!\bigg)\sin(2b\phi_0)\label{eq35}
\\
V_\alpha(\phi_0)\!\!\!&=&\!\!\!c_1\sin^2(b\phi_0)\!+\!c_2\sin^4(b\phi_0)\!+\!c_3\phi \sin(2b\phi_0)\nn\label{eq37}
\een
\ees
where
\ben 
c_1\!\!&=&\!\!a^4\!\left[\!-\frac{320}{81}\!+\!\frac{224b^4}{9}\!+\!\frac{46 b^6}{3}\!+\!b^2 \left(\!\frac{50}{27}\!-\!\frac{45\gamma}{4}\!\right)\!+\!22\gamma\!\right]+\nn
&&+a^2 \left(8+\frac{3 b^2}{2}\right) \beta, \nn
c_2&=&\!\!\!\frac{17a^4}{216}  \left(384 b^4+216 b^6+b^2 (40-243 \gamma )-216 \gamma \right),\nn
c_3&=&\!\!\!\frac{a^2 b}{324}\bigg[\!2a^2\bigg(\!320\!-\!1152 b^3\!+\!576 b^4\!-\!864 b^5\!+\!216 b^6\!+\nn
&&-\!b^2(120-729\gamma)\!-\!1944\gamma\!\bigg)\!-\!81\left(\!8\!+\!6 b\!+\!3 b^2\!\right)\beta \bigg].\nonumber
\een
Also, using the solution \eqref{eq32} we obtain
\ben
\phi(y)\!\!\!&=&\!\!\!\frac1b\bigg[\!1\!+\!\alpha\bigg(\!\frac{40 a^2}{27}\!+\!\frac{8 a^2 b^4}{3}\!-\!9 a^2 \gamma \!-\!\frac{3 \beta }{2}\bigg)\!\bigg]\times\nn
&&\!\!\!\times \arcsin\left[\tanh(a b^2 y)\right]\!\!-\nn
&&\!\!\!-\frac{2\alpha a^2}{b}\bigg(\!\frac{20}{27}\!+\!\frac{64 b^2}{9}\!+\!4 b^4\!-\!\frac{9 \gamma }{2}\!\bigg) \frac{\tanh(a b^2 y)}{\cosh(a b^2 y)}.\nn\label{eq36}
\een

\begin{figure}[!ht]
\begin{center}
\includegraphics[scale=0.56]{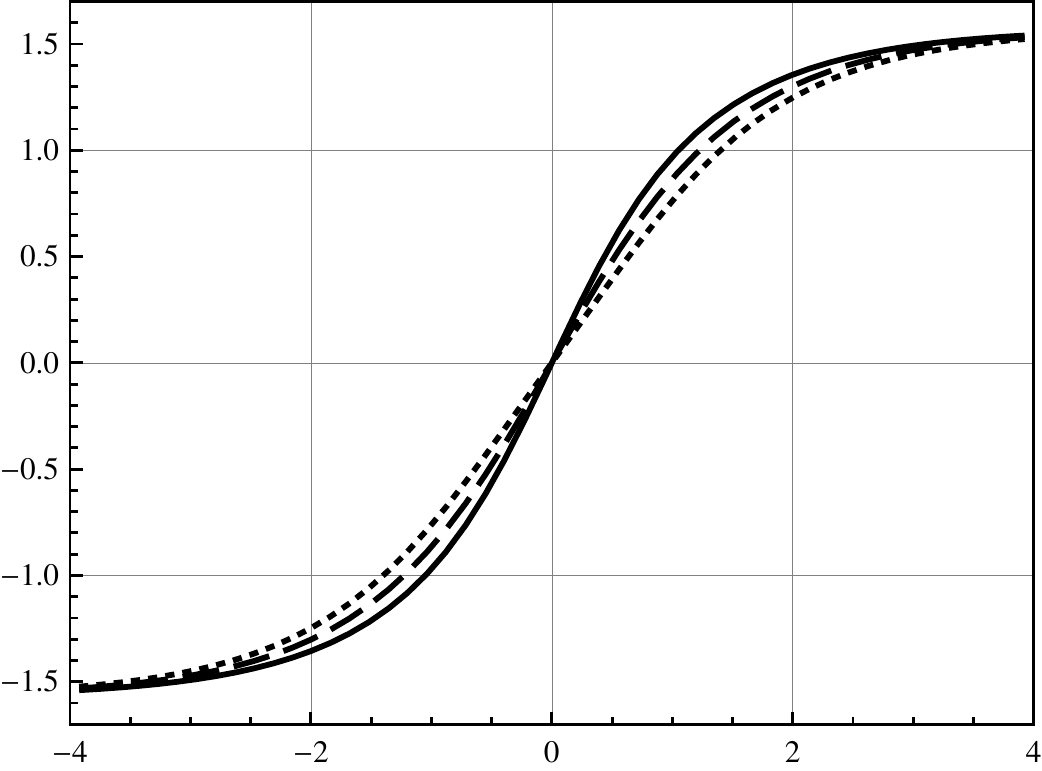}
\includegraphics[scale=0.56]{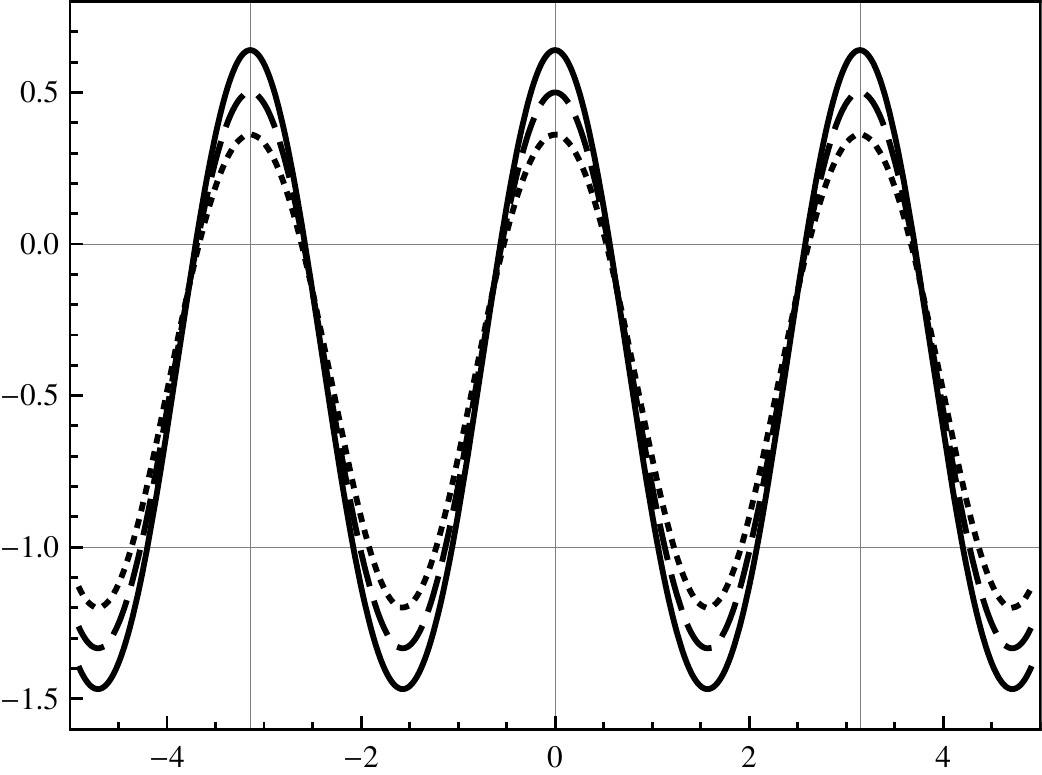}
\end{center}
\caption{\small{Upper panel: solution \eqref{eq36} for $a=b=\beta=1$ and $\gamma=143/486$, for $\alpha=-0.01$ (solid line), $\alpha=0$ (dashed line) and $\alpha=0.01$ (dotted line).
Lower panel: potential \eqref{eq38} for the same values of $a, b, \beta, \gamma$ and $\alpha$.}}\label{figu7}
\end{figure}

The asymptotic valor of $\phi$ is $\phi(y\!\!\rightarrow\! \!\pm\infty)\!=\!\pm\bar\phi$ where
\begin{equation}
\bar\phi=\frac{\pi}{2b}\!+\!\frac{\alpha\pi}{108b}\bigg(80 a^2 \!+\!144 a^2 b^4\!-\!486 a^2 \gamma\!-\! 81 \beta\bigg).
\end{equation}
We see that if $\beta\!=\!(2 a^2/81)\times \left(40\!+\!72 b^4\!-\!243 \lambda \right)$ we have that $\phi(y\!\rightarrow\!\pm\infty )\!=\!\phi_0(y\!\rightarrow\!\pm\infty )$ . In Fig.~\ref{figu7}, in the upper panel  we depict the solution \eqref{eq36} for $a\!=\!b\!=\!\beta\!=\!1$, $\gamma\!=\!143/486$ and some values of $\alpha$. Also, the potential has the form
\ben
V(\phi)\!\!&=&\!\!\!\frac{a^2 b^2}{2}-\frac{1}{6} a^2 \left(8+3 b^2\right) \sin^2(b\phi)+\nn
&+&\!\!\!\alpha\!\bigg[c_1\!\sin^2(b\phi)\!+\!c_2\sin^4(b\phi)\!+\!c_3\phi \sin(2b\phi)\bigg],\label{eq38}
\een
which is depicted in Fig.~\ref{figu7}, in the lower panel. We see that
\begin{equation}
V(\phi\!\to\!\bar\phi)=-\frac{4 a^2}{3}+8\alpha a^2\left[\beta\!+\!4a^2 \bigg(\lambda-\frac{10}{81}\bigg) \right],
\end{equation}
and $V_\phi(\phi\!\to\!\bar\phi)=0$.

%%%%%%%%%%%%%%%%%%%%%%%%%%%%
%%%%%%%%%%%%%%%%%%%%%%%%%%%%

Furthermore, the warp function is
\ben
A(\phi_0)&=&\frac{2 a^2 \alpha \gamma}{b^2}\cos(2 b \phi_0)+\frac{2}{3 b^2}\left[1-3\alpha(\beta+4\gamma a^2) \right]\times\nn
&&\times\ln\left[\cos(b \phi_0)\right]-\frac{2 a^2 \alpha  \gamma }{b^2},
\een
which obeys $A(0)=0$. Using \eqref{eq32} we have
\ben \label{eq33}
A(y)&=&- \frac{4\alpha\gamma a^2}{b^2}+ \frac{4\alpha\gamma a^2}{b^2}S^2+\nn
&&+\frac{2}{3b^2}\left[1 -3\alpha(\beta+4\gamma a^2)\right] \ln (S).
\een
where $S=\sech(a b^2 y)$.

%%%%%%%%%%%%%%%%%%%%%%%%%%%%
\begin{figure}[!ht]
\begin{center}
\includegraphics[scale=0.56]{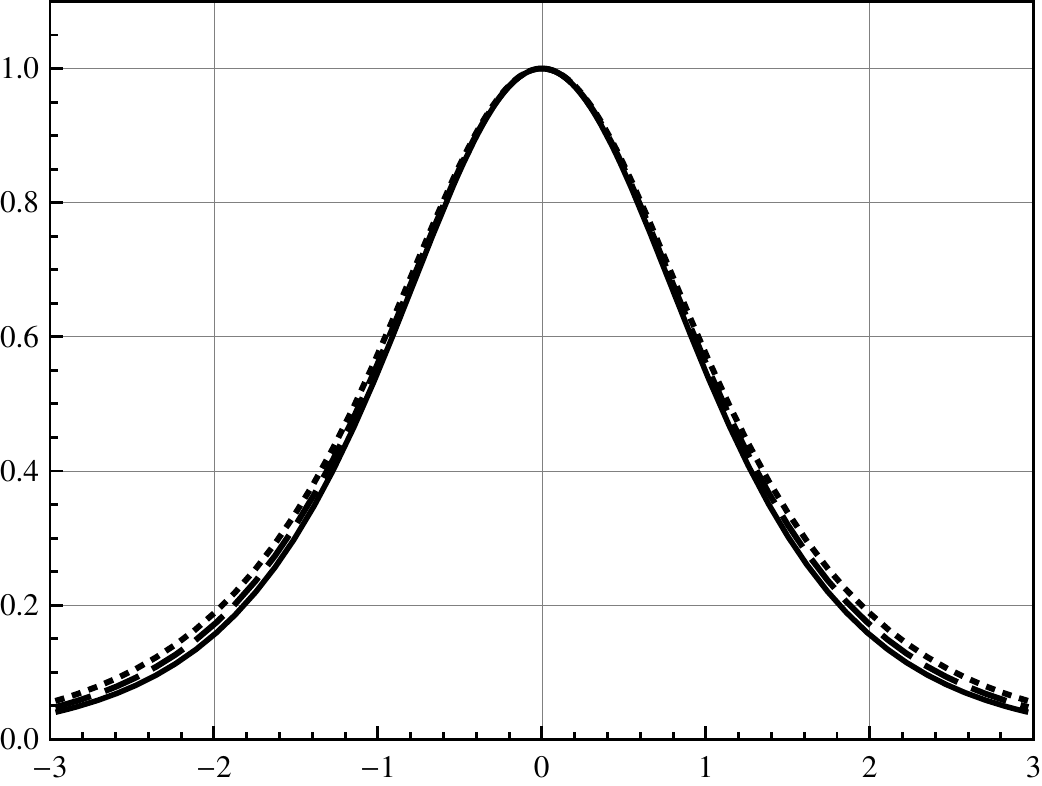}
\includegraphics[scale=0.56]{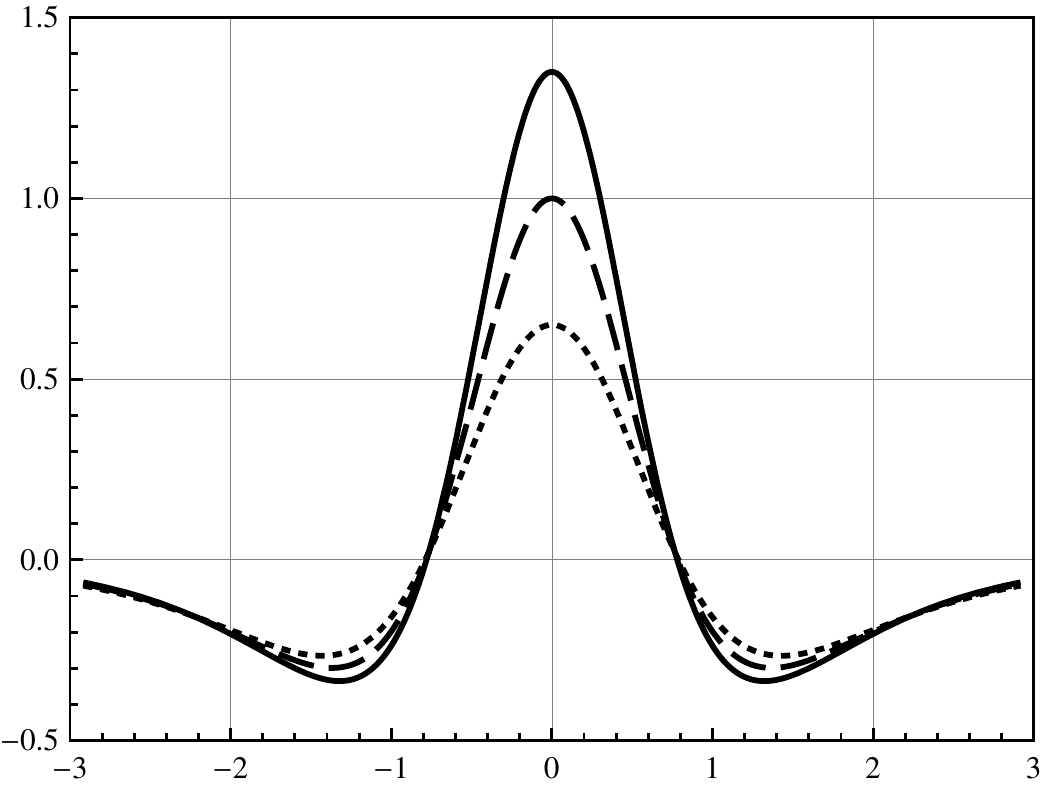}
\end{center}
\caption{\small{Upper panel: warp Factor \eqref{eq33} for $a=b=\beta=1$ and $\gamma=143/486$,  for $\alpha=-0.01$ (solid line), $\alpha=0$ (dashed line) and $\alpha=0.01$ (dotted line). Lower panel: energy density \eqref{eq34} for the same values of $a, b, \beta, \gamma$ and $\alpha$.}}\label{figu8}
\end{figure}
%%%%%%%%%%%%%%%%%%%%%%%%%%%%

The energy density is
\ben\label{eq34}
\rho(y)\!\!\!&=&\!\!\!\frac{1}{3} a^2 \!\left[2\!+\!3 b^2\!-\!2 ~\!\sech^{\!\!-1}\!(2 a b^2 y)\right]\!\!S^{\frac{6 b^2+4}{3 b^2}}-\nn
& &\!\!\!-\frac{a^2\alpha}{162 b^2} S^{\frac{4}{3 b^2}}\bigg[d_1S^4+d_2 S \tanh(ab^2 y)\times\nn
&&\!\!\!\times\arcsin\left[\tanh(ab^2y)\right]\!+\!\sum_{i=1}^{2}\!\bigg(\!g_i\!+\!h_i\ln(S)\!\bigg)\!S^{2i-2}\bigg],\nn
\een
where
\ben
d_1\!\!&=&\!\!18 a^2\! \left(40 b^4\!+\!384 b^6\!+\!216 b^8\!-\!96 \gamma \!-\!144 b^2 \gamma \!-\!243 b^4 \gamma \right) \nn
d_{2}\!\!&=&\!\!b^2 \left(8\!+\!3 b^2\right) \left(81\beta \!-\!2 a^2 \left(40\!+\!72 b^4\!-\!243 \gamma \right)\right)\nn
g_1\!\!&=&\!\!640 a^2 b^2-1728 a^2 \gamma -5184 a^2 b^2 \gamma -1296 b^2 \beta\nn
g_{2}\!\!&=&\!\!-3456 a^2 \gamma -864 \beta  \nn
h_1\!\!&=&\!\!a^2\! \left[27\gamma(8\!+\!9 b^2)^2\!-\!8b^2 (40\!+\!9b^2(5\!+\!24b^2\!+\!15b^4))\right]\!+\nn
&&\!\!+\;81b^2\beta (8\!+\!3 b^2)\nn
h_{2}\!\!&=&\!\!216 \left(4+3 b^2\right) \left(4 a^2 \gamma +\beta \right)\nonumber
\een

In Fig.~\ref{figu8}, in the upper panel we depict the warp fact $e^{2A(y)}$, for some values of the parameters. For this model we must have $\alpha<[3 \left(4 a^2 \gamma +\beta \right)]^{-1}$, to make $e^{2A(y)}>0$. In Fig.~\ref{figu8}, in the lower panel we depict the behavior of the energy density \eqref{eq34}, for the same values of parameters.

The scalar curvature becomes
\ben\nonumber
R(y)\!\!\!&=&\!\!\!\frac{16a^2}{9}\!\left[5 \!-\! \left(5\!+\!3 b^2\right)S^2\right]\!-\!\frac{160}{3}\alpha a^2\!\! \left(4 a^2 \gamma \!+\!\beta \right)\!-\nn
&&\!\!\!-\frac{64}{3}\alpha\gamma a^4 \left(10+9 b^2\right)S^4+\nn
& &\!\!\!+\frac{16a^2\alpha}{3}\!\bigg(\!80 a^2 \gamma\!+\!36 a^2 b^2 \gamma \!+\!10 \beta \!+\!3 b^2 \beta\! \bigg)S^2.\nn
\een
It goes asymptotically to the constant value $R(y\!\to\!\pm \infty)= (80a^2/9)\times\left[1-6\alpha (4 a^2 \gamma +\beta )\right]$. Also, at the origin it gives $R(0)\!=\!(16a^2 b^2/3) \times (3\alpha \beta -1)$.

%%%%%%%%%%%%%%%%%%%%%%%%%%%%
\section{Comments and conclusions}

In this work we succeeded to find exact and approximated solutions for the warp factor, the scalar field $\phi$ and energy density in models of $F(R)$ brane with a non-constant curvature. We used $F(R)=R+\alpha R^2$ to solve the equations of motion, and we studied models with the potential for the scalar field engendering polynomial or nonpolynomial interactions. For several distinct examples, we showed that the warp factor is indeed a well-behaved function, the scalar field displays kinklike behavior, and the profile of the energy density appears as expected.

Interestingly, we have found a situation where brane splitting behavior may appear, induced by the parameter $\alpha$, which controls the way the generalized gravity enters the game. This effect is different from the brane splitting behavior found in \cite{S1,S2}, where thermal or 2-kink effects may induce the splitting, in models with standard gravity. In this work, the splitting is directly related to the parameter that controls deviation from standard gravity.

We have investigated how the addition of higher order power in the curvature may contribute to the splitting of the brane. In the case with $F(R)=R+ \alpha R^n$, for $n=3,4,...$, we have checked up to $n=5$, that the brane splitting effect works as in the case studied before, for $n=2$. The results suggest that the brane splitting is a generic effect, for the above polynomial modification of the standard gravity. Further details of the calculations will be given elsewhere.

The authors would like to thank CAPES and CNPq for partial financial support. The work by A. Yu. P. has been supported by the CNPq project 303438/2012-6.

%%%%%%%%%%%%%%%%%%%%%%%%%%%%


\begin{thebibliography}{99}

\bibitem{RS} 
 L.~Randall and R.~Sundrum, Phys.\ Rev.\ Lett.\  {\bf 83}, 4690 (1999).

\bibitem{Nima} N. Arkani-Hamed, S. Dimopoulos, and G. Dvali, Phys. Lett. B {\bf429}, (1998); I. Antoniadis, N.
Arkani-Hamed, S. Dimopoulos, and G. Dvali, Phys. Lett. B {\bf436}, 257 (1998).

\bibitem{GW} W. D. Goldberger and M. B. Wise, Phys. Rev. Lett. {\bf83}, 4922 (1999).

\bb{AH}N. Arkani-Hamed, S. Dimopoulos, G. Dvali, and N. Kaloper, Phys. Rev. Lett. {\bf84}, 586 (2000).

\bibitem{F} O. DeWolfe, D.Z. Freedman, S. Gubser, and A. Karch, Phys. Rev. D {\bf62}, 046008 (2000); C. Csaki,
J. Erlich, T. Hollowood, and Y. Shirman, Nucl. Phys. B {\bf581}, 309 (2000); C. Csaki, J. Erlich, G.
Grojean, and T. Hollowood, Nucl. Phys. B {\bf584}, 359 (2000).
\bibitem{G} M. Gremm,  Phys.\ Lett.\ B {\bf 478}, 434 (2000).

\bb{S1}A. Campos, Phys. Rev. Lett. {\bf88}, 141602 (2002).
\bb{S2}D. Bazeia, C. Furtado, and A.R. Gomes, JCAP {\bf0402}, 002 (2004);

\bibitem{D} G. Dvali, G. Gabadadze, and M. Porrati, Phys. Lett. B {\bf485}, 208 (2000).
A. Karch and L. Randall, JHEP {\bf0105}, 008 (2001);
M. Porrati, Phys. Lett. B {\bf498}, 92 (2001); F.A. Brito, M. Cvetic, and S.-C. Yoon, Phys. Rev. D {\bf64}, 064021 (2001);
M. Cvetic and N.D. Lambert, Phys. Lett. B {\bf540}, 301 (2002); 
A. Melfo, N. Pantoja, and A. Skirzewski, Phys. Rev. D {\bf67}, 105003 (2003); D. Bazeia, F.A. Brito, and J.R. Nascimento,
Phys. Rev. D {\bf68}, 085007 (2003); O. Castillo-Felisola, A. Melfo, N. Pantoja, and A. Ramirez, Phys. Rev. D {\bf70}, 104029 (2004);
K. Takahashi and T. Shiromizu, Phys. Rev. D {\bf70}, 103507 (2004); A. Celi et al. Phys. Rev. D {\bf71}, 045009 (2005);
A. Celi, JHEP {\bf0702}, 078 (2007); A. Ceresole and G. Dall'Agata, JHEP {\bf0703}, 110 (2007).

\bibitem{C} 
  C.~Csaki, {\it TASI Lectures on Extra Dimensions and Branes,} [hep-ph/0404096].

\bibitem{B} 
 D. Bazeia and A.R. Gomes, JHEP {\bf0405}, 012 (2004); D. Bazeia, F. Brito, and L. Losano, JHEP {\bf0611}, 064 (2006).
 V.I. Afonso, D. Bazeia, and L. Losano, Phys. Lett. B {\bf634}, 526 (2006). 

\bibitem{AB} 
  V.I. Afonso, D. Bazeia, R. Menezes and A.Y. Petrov, Phys. Lett. B {\bf 658}, 71 (2007).

\bibitem{BMP} D. Bazeia, R. Menezes, A.Yu. Petrov, and A.J. da Silva, Phys. Lett. B {\bf726}, 523 (2013).

\bb{mg}
 S.M. Carroll et al., Phys. Rev. D {\bf71}, 063513 (2005); G. Cognola et al., JCAP {\bf0502}, 010 (2005);
M. Amarzguioui, O. Elgaroy, D.F. Mota, and T. Multamaki, Astron. Astrophys. {\bf454}, 707 (2006);
S. Capozziello, S. Noriji, S.D. Odintsov, and A. Troisi, Phys. Lett. B {\bf 639}, 135 (2006); S. Noriji and S.D. Odintsov, Phys. Rev. D {\bf74}, 086005 (2006);
L. Amendola, D. Polarski, and S. Tsujikawa, Phys. Rev. Lett. {\bf98}, 131302 (2007); S. Noriji and S.
Y.-S. Song, W. Hu, and I. Sawicki, Phys. Rev. D {\bf75}, 044004 (2007);
L. Amendola, R. Gannouji, D. Polarski, and S. Tsujikawa, Phys. Rev. D {\bf75}, 083504 (2007);
I. Sawicki and W. Hu, Phys. Rev. D {\bf75}, 127502 (2007);
M. Cvetic and M. Robnik, Phys. Rev. D {\bf77}, 124003 (2008); 
D. Bazeia, A. R. Gomes, and L. Losano, Int. J. Mod. Phys. A {\bf24}, 1135 (2009);
Y.-X. Liu, Z.-H. Zhao, S.-W. Wei, and Y.-S. Duan, JCAP {\bf02}, 003 (2009); 
C.A.S. Almeida, M. M. Ferreira Jr., A. R. Gomes, and R. Casana, Phys. Rev. D {\bf79},125022 (2009);
Y. Zhong, Y.-X. Liu, and K. Yang, Phys. Lett. B {\bf699}, 398 (2011); Y.-X. Liu, Y. Zhong, and Z.-H. Li, JHEP {\bf1106}, 135 (2011);
A. Ahmed and B. Grzadkowski, JHEP {\bf1301}, 177 (2013).

\bb{FO}M. Cvetic, S. Griffies and S.-J. Rey, Nucl. Phys. B {\bf 381}, 301 (1992);
D. Bazeia, L. Losano, and R. Menezes, Phys. Lett. B {\bf668}, 246 (2008); 
D. Bazeia, A.R. Gomes, L. Losano, R. Menezes, Phys. Lett. B {\bf671}, 402 (2009). 

\bibitem{ABLM} C.A.G. Almeida, D. Bazeia, L. Losano, and R. Menezes, Phys. Rev. D {\bf88}, 025007 (2013). 
\end{thebibliography}
\end{document}